\documentclass[10pt]{article}
\usepackage{amsmath}
\usepackage{cite} 
\usepackage{indentfirst}
\usepackage{latexsym}
\usepackage{pstricks,pst-node}

\long\def\ca#1\cb{} 

\newcommand{\ad}{^\dagger }

\newcommand{\AND}{{\small AND}}

\newcommand{\becs}{\begin{cases}}
\newcommand{\bem}{\begin{matrix}}

\newcommand{\dya}[1]{|#1\rangle\langle#1|}

\newcommand{\encs}{\end{cases}}
\newcommand{\enm}{\end{matrix}}

\newcommand{\inpd}[2]{\langle#1|#2\rangle }
\newcommand{\ket}[1]{|#1\rangle }

\newcommand{\lra}{\leftrightarrow }

\newcommand{\msk}{\medskip }

\newcommand{\mte}[2]{\langle#1|#2|#1\rangle }

\newcommand{\NOT}{{\small NOT}}
\newcommand{\od}{\odot }
\newcommand{\OR}{{\small OR}}
\newcommand{\ot}{\otimes }

\newcommand{\ra}{\rightarrow }

\newcommand{\vbB}{\boldsymbol{\mid}}

\newcommand{\FC}{{\mathcal F}}

\newcommand{\rB}{\textbf{r}}

\newcommand{\al}{\alpha }
\newcommand{\bt}{\beta }
\newcommand{\gm}{\gamma }
\newcommand{\Gm}{\Gamma }
\newcommand{\dl}{\delta }

\newcommand{\om}{\omega }

\def\outl#1{\par{\medskip\noindent\hspace*{.2cm}\bf
      \mathversion{bold}#1\mathversion{normal}\smallskip} }
 \def\xa{} \def\xb{}  

 \def\outl#1{}  \def\xa{} \def\xb{}  

\ca
 \def\outl#1{\par{\medskip\noindent\hspace*{.5cm}\bf
      \mathversion{bold}#1\mathversion{normal}\smallskip} }
 \long\def\xa#1\xb{}
 
\cb

\begin{document} 
\ca 
okon/sok04.tex\ \ %

\begin{center}
{\large Consistent Quantum Measurements}\\
Robert B. Griffiths\\
Version of 27 July 2015
\end{center}

\msk
\cb

\title{Consistent Quantum Measurements}

\author{Robert B. Griffiths
\thanks{Electronic mail: rgrif@cmu.edu}\\ 
Department of Physics,
Carnegie-Mellon University,\\
Pittsburgh, PA 15213, USA}
\date{Version of 27 July 2015}
\maketitle  
\ca
\centerline{Robert B. Griffiths}
\centerline{Physics Department}
\centerline{Carnegie-Mellon University}
\vspace{.2cm}
\cb

\xb

\xa
\begin{abstract}
  In response to recent criticisms by Okon and Sudarsky, various aspects of the
  consistent histories (CH) resolution of the quantum measurement problem(s)
  are discussed using a simple Stern-Gerlach device, and compared with the
  alternative approaches to the measurement problem provided by spontaneous
  localization (GRW), Bohmian mechanics, many worlds, and standard (textbook)
  quantum mechanics.  Among these CH is unique in solving the second
  measurement problem: inferring from the measurement outcome a property of the
  measured system at a time before the measurement took place, as is done
  routinely by experimental physicists.  The main respect in which CH differs
  from other quantum interpretations is in allowing multiple stochastic
  descriptions of a given measurement situation, from which one (or more) can
  be selected on the basis of its utility.  This requires abandoning a
  principle (termed unicity), central to classical physics, that at any instant
  of time there is only a single correct description of the world.
\end{abstract} \xb

\tableofcontents

\section{Introduction}
\label{sct1}

\xb
\outl{Critique of CH by OS is motivation for paper
}\xa

\xb
\outl{CH the most advanced Qm interpretation; resolves all paradoxes 
}\xa

\xb
\outl{CH results currently available; not just in future publications
}\xa

The immediate motivation for this paper comes from criticisms by Okon and
Sudarsky \cite{OkSd14b}, recently published in this journal, of the
\emph{consistent histories} (CH) interpretation of quantum mechanics.  These
authors claim that CH does not provide a satisfactory resolution of the quantum
measurement problem.  Such criticism deserves to be taken seriously, for the CH
approach claims to resolve \emph{all} the standard problems of quantum
interpretation which form the bread and butter of quantum foundations research:
it is local \cite{Grff11}, so there are no conflicts with special relativity;
it is noncontextual \cite{Grff13b}, in contrast to hidden variables
interpretations; it resolves the EPR, BKS, Hardy, three boxes, etc., etc.\
paradoxes, see Chs.~19-25 of \cite{Grff02c}.  And while it may be defective,
its (purported) solutions to the full gamut of quantum conceptual difficulties
have been published in detail and are available right now for critical
inspection, not just as promissory notes for some future time.  Thus the
Okon and Sudarsky criticisms, while based (we believe) on an imperfect
understanding of the CH approach, are dealing with important issues that need
to be discussed.

\xb \outl{CH: measurement not a basic postulate, just a physical process.  So
  no measurement problem.  }\xa

\xb
\outl{CH resolves 2d measurement problem
}\xa

Of particular significance is the fact that the CH approach does \emph{not}
include \emph{any} reference to measurements among its basic principles for
interpreting quantum mechanics. Measurements are simply treated as a particular
type of physical process to which the same quantum principles apply as to any
other physical process.  When understood in this way quantum mechanics no
longer has a \emph{measurement problem} as that term is generally used in
quantum foundations: a conflict between unitary time development of a combined
system plus measuring device and a macroscopic outcome or ``pointer position.''
Not only so, in addition CH shows how the outcome of a measurement can be
shown to reveal the presence of a microscopic quantum property possessed by the
measured system just \emph{before} the measurement took place, in accordance
with the belief, common among experimental physicists, that the apparatus they
have built performs the function for which it was constructed.  This
\emph{second} measurement problem has received far too little attention in the
quantum foundations literature, and resolving it is no less important than the
first problem if the entire measuring process is to be understood in fully
quantum-mechanical terms.

\xb \outl{Particular example will show how CH, other approaches (BM,GRW,SQM,MW)
  handle measurements }\xa

\xb
\outl{References to material giving intro to CH
}\xa
 
Rather than an abstract discussion, the present paper examines a particular
measurement scenario, using it as an example of the application of CH
principles, and also a basis for comparison with some other interpretations of
quantum mechanics mentioned in \cite{OkSd14b}. These include the
\emph{spontaneous localization} approach developed by Ghirardi et al.\ and
Pearle, see \cite{GhRW85,GhRW86,Prl89,Frgg09,Ghrr11}, often abbreviated as GRW
(the initials of the authors of \cite{GhRW85}), and the pilot wave approach of
de Broglie and Bohm, which we shall refer to as \emph{Bohmian mechanics}
\cite{dBrg27,Bhm52a,Hlln93,Glds12}.  Textbook or \emph{standard quantum
  mechanics} and the \emph{many worlds} interpretation of Everett and his
successors,\cite{Evrt57,DWGr73,SBKW10} also enter the discussion from time to
time.  Since details of the CH approach are readily available in the
literature, e.g., \cite{Grff02c,Grff09b,Hrtl11,Grff13,Grff14,Grff14b}, only
those aspects needed to make the discussion reasonably self-contained are
included in this paper.

\xb \outl{CH: QM gives difficulty by abandoning unicity; $\lra$ abandoning
  stationary earth }\xa

Our aim is to present and discuss as clearly as possible the central features
of the CH approach that have given rise to the criticisms in \cite{OkSd14b},
and which are undoubtedly shared by other critics, e.g.,
\cite{Knt98,BsGh00,Prl05,Mrmn13}.  Of particular importance is the fact that CH
abandons a principle, here called \emph{unicity}, which is deeply embedded in
both conventional and scientific thought, and is taken for granted in classical
physics. It is the idea that at any instant of time there is precisely one
exact description of the state of the world which is true.  If the CH
understanding is correct, quantum mechanics has made unicity obsolete in
somewhat the same way as modern astronomy has replaced an unmovable earth at
the center of the universe with our current understanding of the solar system,
and ignoring this feature of the quantum world is what has given rise to
so many conceptual difficulties.

\xb
\outl{Summary of following sections
}\xa

The contents of the remainder of the paper are as follows. The measurement
problem(s) of quantum foundations are discussed in general terms in
Sec.~\ref{sct2}, followed in Sec.~\ref{sct3} by a specific measurement model, a
modernized version of the famous experiment of Stern and Gerlach
\cite{Strn21,GrSt22b}.  Its description in CH terms begins in Sec.~\ref{sct4}
with a discussion of the first measurement problem, whose solution is compared
with some other approaches in Sec.~\ref{sbct4.2}.  The CH solution to the
second measurement problem is the subject of Sec.~\ref{sct5}, and it is
compared with standard quantum mechanics, spontaneous localization, many
worlds, and Bohmian mechanics in Sec.~\ref{sct6}.  Our response to the specific
criticisms of Okon and Sudarsky occupies Sec.~\ref{sct7}.  The concluding
Sec.~\ref{sct8} is a brief summary of the whole paper.

\section{The Quantum Measurement Problem}
\label{sct2}

\xb \outl{Physics an experimental science; measurements important, but CP had
  no measurement problem }\xa

\xb
\outl{Measurement problem: describe what goes on in Qml terms
}\xa

\xb \outl{Microscopic system, macroscopic outcome, processes connecting them;
  discussed without arm waving and evasion as criticized by Bell }\xa

\xb
\outl{Would cosmologists be credible if they did not understand operations of
  telescopes?
}\xa

Physics is an experimental science, and measurements and observations play a
central role in testing the empirical contents of its theories.  This was also
the case before the quantum revolution of the twentieth century, and yet
classical physics had no measurement problem.  Why, then, is the measurement
problem considered \emph{the} central issue in quantum foundations, the one
that must be resolved if progress is to be made in this field?  The essence of
the measurement problem is easy to state. If quantum mechanics applies not only
to the microscopic world of nuclei and atoms, but also to macroscopic objects
and things that are even larger---from the quarks to the quasars---then the
measurement process in which an earlier microscopic property is revealed in a
macroscopic outcome should itself be describable, at least in principle, in
fully \emph{quantum mechanical} terms.  Applied equally to the system being
measured and to the macroscopic apparatus, and without the evasion and
equivocation ridiculed by Bell \cite{Bll90}.  It is indeed a scandal that the
quantum physics community has not been able to agree on a solution to this
problem.  Would not the stories told by modern cosmologists be dismissed as
pure fantasy if astronomers did not understand the operation of their
telescopes?

\xb
\outl{First \& second measurement problem defined
}\xa

\xb
\outl{Textbooks confuse preparation and measurement
}\xa

It is useful to separate the general quantum measurement problem into two
parts. The better known \emph{first} measurement problem arises when the
initial state of the measured system---hereafter for convenience thought of as
a particle---is such that the unitary time development resulting from coupling
it to a measurement apparatus results in a superposition of two or more states
in which the apparatus pointer (in the archaic but picturesque language of
quantum foundations) points in different directions.  How is this
``Schr\"odinger cat'' to be interpreted, given that in the laboratory the
pointer always points in a definite direction?  The \emph{second} measurement
problem is to explain how the actual (single) pointer direction is related to
the property of the particle the apparatus was designed to measure, at a time
\emph{before} the measurement took place.  Unfortunately, many textbooks speak
of a ``measurement'' not as revealing a pre-existing property, but as a
correlation between the pointer and the particle \emph{after} the measurement
has taken place.  The latter should be called a \emph{preparation} rather than
a measurement; for a discussion of this from the CH perspective see Sec.~3.5 of
\cite{Grff14} and Sec.~7.3 of \cite{Grff14b}.

\xb
\outl{Textbook QM introduces probabilities thru measurements; unresolved
  measurement problem $\ra$ inconsistent discussion of probabilities.
}\xa

It is perhaps worth mentioning that in textbooks \emph{probabilities} are
introduced in connection with measurements, and not as a separate topic.  As a
consequence the perplexities associated with an unresolved measurement problem
are transferred to an inconsistent discussion of probabilities.  Thus cleaning
up the quantum measurement problem is intimately connected with introducing
probabilities in quantum mechanics in a consistent way, not associated with
measurements, something which is not present in any textbook of which we are
aware.

\section{Stern Gerlach Spin Measurement}
\label{sct3}

\subsection{Description}
\label{sbct3.1}

\begin{figure}[h]
$$
\begin{pspicture}(-2,-1.5)(+4.0,+1.0)
\psset{
arrowsize=0.150 0,linewidth=0.025,dash=0.1 0.1}
\newpsobject{showgrid}{psgrid}{subgriddiv=1,griddots=10,gridlabels=6pt}
\def\lwp{0.035} 
\def\wprad{0.35} 
\def\wpran{-0.35} 
\def\circb{
\pscircle[fillcolor=white,fillstyle=solid]{0.30}}
\def\dput(#1)#2#3{\rput(#1){#2}\rput(#1){#3}}
\def\lwdsh{0.03}
\def\tnline(#1,#2,#3)#4{
\rput(#1,#2){\psline[linestyle=dashed,linewidth=\lwdsh](0.0,-0.3)(0.0,#3)}
\rput(0,-0.7){\rput[B](#1,#2){#4}}}
\psline(0,+1)(0,\wprad)(1.5,\wprad)(1.5,+1)
\psline(0,-1)(0,\wpran)(1.5,\wpran)(1.5,-1)
\rput(0.75,-0.60){Magnet}
\psline[linewidth=\lwp](-2,0)(0.5,0)
\psline[linewidth=\lwp]{->}(-2,0)(-1.2,0)
\psbezier[linewidth=\lwp](0,0)(0.5,0)(1.0,0)(1.5,+0.125)
\psline[linewidth=\lwp](1.5,+.125)(3,+0.5)
\psline[linewidth=\lwp]{->}(1.5,+.125)(1.9,+0.225)
\psbezier[linewidth=\lwp](0,0)(0.5,0)(1.0,0)(1.5,-0.125)
\psline[linewidth=\lwp](1.5,-.125)(3,-0.5)
\psline[linewidth=\lwp]{->}(1.5,-.125)(1.9,-0.225)
\psline(2.925,+0.8)(3.075,+0.2)(3.875,+0.4)(3.725,+1.0)(2.925,+0.8)
\psline(2.925,-0.8)(3.075,-0.2)(3.875,-0.4)(3.725,-1.0)(2.925,-0.8)
\rput(3.4,0.57){$D^a$}
\rput(3.4,-0.55){$D^b$}
\dput(-2,0){\circb}{$\om_0$}
\dput(-0.5,0){\circb}{$\om_1$}
\dput(2.4,0.35){\circb}{$\om_2^a$}
\dput(2.4,-0.35){\circb}{$\om_2^b$}
\tnline(-2,-0.7,.3){$t_0$}
\tnline(-0.5,-0.7,.3){$t_1$}
\tnline(2.4,-0.7,.0){$t_2$}
\tnline(3.6,-0.7,-0.3){$t_3$}
\end{pspicture}
$$
\caption{Stern Gerlach apparatus for measuring spin half
}
\label{fgr1}
\end{figure}
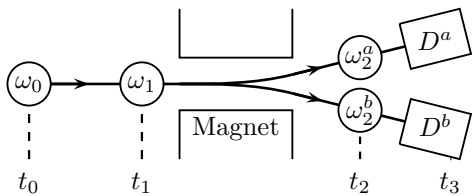

\xb
\outl{Particle, detectors, kets, unitary time development
}\xa

Figure~\ref{fgr1} is a schematic diagram of a Stern Gerlach device to measure
the spin of a spin-half particle.  The particle arrives from the left and its
initial state at time $t_0$ is $\ket{\om_0}\ot\ket{\chi_0}$, where $\ket{\om_0}$
refers to its position, corresponding to a wavepacket
$\om_0(\rB)=\inpd{\rB}{\om_0}$, and $\ket{\chi_0}$ denotes the spin, with
$\ket{z^+}$ and $\ket{z^-}$ the eigenstates of $S_z$.  The unitary time
development of the particle state at successive times $t_0<t_1<t_2$ as it passes
through the magnetic field gradient is given by:
\begin{equation}
 \ket{\om_0}\ot\ket{z^+} \ra \ket{\om_1}\ot\ket{z^+} \ra
 \ket{\om_2^a}\ot\ket{z^+};\quad 
 \ket{\om_0}\ot\ket{z^-} \ra \ket{\om_1}\ot\ket{z^-} \ra
 \ket{\om_2^b}\ot\ket{z^-},
\label{eqn1}
\end{equation}
where $\ket{\om_j}$ gives the (approximate) location of the particle at time
$t_j$.  The trajectories of a particle with $S_z=+1/2$ and one with $S_z=-1/2$
are initially identical, but at time $t_2$ there is a small but macroscopic
separation between the wave packet $\om_2^a(\rB)$, the particle moving upwards
towards detector $D^a$, and $\om_2^b(\rB)$, the particle moving downwards
towards detector $D^b$.  By time $t_3$ the detector $D^a$ will have triggered
if the particle had $S_z=+1/2$, and $D^b$ if the particle had $S_z=-1/2$.  We
assume these detectors are capable of detecting individual atoms, as is
possible nowadays by first ionizing the atom and then using an electron
multiplier to convert the emerging electron into a macroscopic current pulse.

\xb
\outl{Detectors $a$, $b$ assigned kets for ready and triggered states
}\xa

Now let us treat the detectors as quantum objects by assigning kets
$\ket{D^a}$, $\ket{D^b}$ for their initial (ready) states, and $\ket{D^{a*}}$
and $\ket{D^{b*}}$ if the detector has been triggered by absorbing the
particle.  If we assume that at $t_0$ the spin state of the particle is
\begin{equation}
\ket{\chi_0}=\al\ket{z^+}+\bt\ket{z^-},
\label{eqn2}
\end{equation}
the unitary time development of the detectors along with the particle, regarded
as a single closed quantum system, is given by
\begin{equation}
\ket{\Psi_0} \ra \ket{\Psi_1} \ra \ket{\Psi_2} \ra \ket{\Psi_3},
\label{eqn3}
\end{equation}
where 
\begin{align}
\ket{\Psi_j} &= \ket{\om_j}\ot(\al\ket{z^+} + \bt\ket{z^-})\ot 
\ket{D^a}\ot \ket{D^b} \text{ for $j=0,\,1$ };
\label{eqn4}\\
 \ket{\Psi_2} &= (\al\ket{\om_2^a}\ot\ket{z^+}+\bt\ket{\om_2^b}\ot\ket{z^-})
\ot\ket{D^a}\ot \ket{D^b};
\label{eqn5}\\
 \ket{\Psi_3} &= 
\al\ket{D^{a*}}\ot \ket{D^b} + \bt \ket{D^a}\ot \ket{D^{b*}},
\label{eqn6}
\end{align}
(Notice that there is no separate
particle state at $t_3$, as the particle has been absorbed into the detector
and its spin state is no longer relevant.)

\subsection{Discussion}
\label{sbct3.2}

\xb
\outl{1st measurement problem. MQS state for detectors
}\xa

The (first) measurement problem of quantum foundations is apparent in
\eqref{eqn6} in a situation in which both $\al$ and $\bt$ are nonzero, and thus
$\ket{\Psi_3}$ is a superposition of two states which are macroscopically quite
distinct: in one case $D^a$ has triggered, and in the other case $D^b$.  There
seems to be some sort of conflict between unitary time evolution and the
existence of a definite measurement outcome in which either $D^a$ has triggered
(as indicated by some pointer position) and $D^b$ remains in the ready state,
or $D^b$ has triggered and $D^a$ has not.  

\xb
\outl{How to understand MQS state? FAPP approach of SQM
}\xa

\xb
\outl{Decoherence does not solve the problem
}\xa

How is this to be understood? Let us consider some of the better-known
approaches to this problem, starting with standard or textbook quantum
mechanics. Here the usual strategy is to give various reasons why ``for
all practical purposes'' \eqref{eqn6} is to be understood as telling us that
with probability $|\al|^2$ detector $D^a$ has been triggered, and $D^b$ with
probability $|\bt|^2$.  For a detailed critique of this approach, see
\cite{Bll90}.  Few physicists working in the area of quantum foundations
consider this solution satisfactory, even though it employs the sort of
reasoning which is known to agree with laboratory experiments.  Nowadays
decoherence due to an environment is sometimes invoked in order to justify the
this approach, but by itself it does not solve the problem, as the environment
can always be included in the detector states $\ket{D^a}$, $\ket{D^{a*}}$
$\ket{D^b}$, and $\ket{D^{b*}}$; see, for example \cite{Adlr03}.

\xb
\outl{GRW, BM, MW approaches to the problem
}\xa

In the spontaneous localization interpretation the unitary dynamics leading
from $t_2$ to $t_3$, i.e., from \eqref{eqn5} to \eqref{eqn6}, should be
replaced by a stochastic dynamics in which, because detectors consist of a
large number of particles, with some probability $\ket{D^{a*}}\ot \ket{D^b}$,
or at least some ket that represents the same macroscopic physics, will be the
correct quantum state at time $t_3$, and with some probability it will be
$\ket{D^a}\ot \ket{D^{b*}}$.  In Bohmian mechanics a collection of
``hidden variables'', particle positions, are added to the Hilbert space
description, and thus whereas $\ket{\Psi_3}$ is the correct wave function at
time $t_3$, the particles which make up the detectors will either correspond to
the situation $\ket{D^{a*}}\ot \ket{D^b}$ or to $\ket{D^a}\ot \ket{D^{b*}}$,
but not both; the part that does not correspond to the actual particle
positions is considered an ``empty wave.''  In the many worlds
interpretation $\ket{\Psi_3}$ is the correct quantum state, but the two parts
of it refer to what are, in effect, two different and unconnected
``universes.''

\xb
\outl{2d measurement problem: outcome implies what about prior location
  or spin of particle?
}\xa

Once the first measurement problem has been solved and good reasons given why
at the end of the experiment the pointer is in a well-defined position, which
is to say one of the detectors in Fig.~\ref{fgr1} has detected the particle and
the other has not, the second measurement problem remains: what can we conclude
about the quantum state of the particle \emph{before} it was measured?  In
particular, if the final state is $\ket{D^{a*}}$, can we conclude that the
particle's location corresponded to $\ket{\om_2^a}$ at time $t_2$, rather than
$\ket{\om_2^b}$?  Or that at $t_1$ the spin state was $\ket{z^+}$ and not
$\ket{z^-}$?  In the world of particle physics experiments, inferences of this
sort are made all the time, and no interpretation of quantum mechanics which
fails to provide a basis for them can be considered a satisfactory explanation
of why quantum theory agrees with experiment.

\section{Consistent Histories and the First Measurement Problem}
\label{sct4}

\subsection{Quantum properties}
\label{sbct4.1}

\xb \outl{Qm properties $\lra$ Hilbert subspaces. Examples: spin 1/2; particle
  in $x_1\leq x \leq x_2$ }\xa

Before discussing how the situation in Fig.~\ref{fgr1} is analyzed in the
consistent histories (CH) approach it is useful to briefly review some of its
key ideas.  The first is that a \emph{physical property} is always represented
by a linear \emph{subspace} of the Hilbert space used to describe the quantum
system; equivalently, by the projector (orthogonal projection operator) onto
this subspace.  As an example, for a spin half particle $[z^+] = \dya{z^+}$ is
the projector onto the one-dimensional subspace or ray consisting of all
multiples of $\ket{z^+}$, and corresponds to the property $S_z=+1/2$ in units
of $\hbar$. (Here and later we use $[\psi]$ as an abbreviation for the
projector $\dya{\psi}$.) For a spinless particle in one dimension the subspace
spanned by all wave packets $\psi(x)$ which are zero outside the interval
$x_1\leq x \leq x_2$ corresponds to the property that the particle is located
between $x_1$ and $x_2$.

\xb
\outl{Analogy: Cl indicators. Negation, conjunction, disjunction
}\xa

A projector is the quantum analog of an indicator function $P(\gm)$ for a
classical property: $P(\gm)$ is 1 at all points $\gm$ in the phase space $\Gm$
where the property is true, and 0 elsewhere. Indicator functions provide a
convenient algebra for discussing classical properties.  Thus if $I$ is the
identity, the function taking the value 1 for all $\gm$, the negation of a
property $P$, ``\NOT\ $P$'', has an indicator $I-P$; while if $P$ and $Q$
represent two properties, $PQ$ is the indicator for the conjunction ``$P$ \AND\
$Q$'', and $P+Q-PQ$ the indicator for the disjunction ``$P$ \OR\ $Q$''.
Replacing classical indicators with quantum projectors leads to identical
expressions for the algebra of quantum properties as long as the projectors $P$
and $Q$ are \emph{compatible}, meaning they commute with each other, in which
case $PQ=QP$ is again a projector and can be interpreted as the property ``$P$
\AND\ $Q$''.

\xb \outl{Approaches to problem of incompatible projectors: Ignore it; Qm
  logic; CH: conjunction not defined }\xa

However, when $P$ and $Q$ do not commute, the products $PQ$ and $QP$ are not
projectors, so neither can represent the conjunction.  There are at least three
approaches to this difficulty.  One is to ignore it, a route leading to many
quantum paradoxes.  Another is to use the system of quantum logic initiated by
Birkhoff and von Neumann \cite{BrvN36}, in which the conjunction of two
properties is represented by the intersection of the corresponding subspaces,
and the rules of propositional logic must be modified; see Sec.~4.6 of
\cite{Grff02c} for a simple example.  A third approach is that of CH: the
conjunction of properties represented by two incompatible projectors is
undefined, ``meaningless,'' and likewise the disjunction.

\xb
\outl{SFR $\lra$ PDI = Qm sample space or framework. Assigning probabilities
}\xa

This leads to the \emph{single framework rule} (SFR), a central principle of
CH, and one that is often misunderstood.  It says that any valid quantum
description of a physical system, i.e., any way of assigning it properties
which may be either true or false, should be based upon a \emph{projective
  decomposition of the identity} (PDI), a collection $\{P_j\}$ of
mutually-orthogonal nonzero projectors that sum to the identity:
\begin{equation} I = \sum_j P_j;\quad P_j = P_j\ad = P_j^2; \quad P_j P_k =
\dl_{jk} P_j.
\label{eqn7}
\end{equation} 
The orthogonality of the projectors ensures that the properties they represent
are mutually exclusive alternatives: if one is true the others must be false.
That they sum to the identity means that one (and only one) must be true.
Thus these properties form a \emph{sample space} as that term is used in
probability theory. They, along with the projectors formed by taking sums of
two or more of the $P_j$'s, constitute an \emph{event algebra}, which in the
histories approach is called a \emph{framework} (a term which is also used for
the sample space).  Probabilities can then be associated with events in the
usual way: given a collection $\{p_j\}$ of nonnegative numbers summing to 1,
the probability of $P_j$ is $p_j$, and of $P_2+P_3$ is $p_2+p_3$, etc.  In
situations in which a single property $P$ is under discussion and a framework
has not (yet) been defined, the implicit framework is $\{P, I-P\}$.

\xb
\outl{Spin half. ``$S_x=1/2$ AND/OR $S_z=1/2$'' makes no sense.
}\xa

In the case of a spin-half particle the PDI associated with $S_x$ is $I = [x^+]
+ [x^-]$, and that associated with $S_z$ is $I=[z^+] + [z^-]$.  However $[x^+]$
and $[z^+]$ do not commute, so a spin-half particle can never be said to be in
a state with both $S_z=+1/2$ and $S_x=+1/2$.  This is consistent with the
observation that every ray in this two-dimensional Hilbert space is associated
with $S_w=+1/2$ for some direction $w$ in space; there are no possibilities
left over to express ``$S_x=+1/2$ \AND\ $S_z=+1/2$''; there is no room in the
Hilbert space for such a possibility.  (For the quantum logic approach to this
situation see Sec.~4.6 of \cite{Grff02c}.)  Similarly ``$S_x=+1/2$ \OR\ $S_z=+1/2$''
makes no sense, because the disjunction must refer to projectors that commute
with each other; otherwise they are not associated with a single PDI, and no
logical comparison is possible.

\xb
\outl{SFR allows CONSTRUCTING incompatible frameworks, forbids COMBINING them
}\xa

\xb
\outl{Two PDIs compatible iff all projectors commute; otherwise incompatible
}\xa

\xb
\outl{Common refinement of compatible frameworks
}\xa

\xb
\outl{Qm reasoning rules in Ch. 16 of CQT
}\xa

It is important to note that the SFR is not a prohibition against constructing
many incompatible descriptions using various incompatible frameworks. The
theoretical physicist has perfect liberty to make up as many as desired. What
the SFR prohibits is \emph{combining} incompatible frameworks to make up a
single description or carry out a single logical argument in the manner which,
for example, leads to a Kochen-Specker paradox (see the discussion in Ch.~22 of
\cite{Grff02c}).  Given two PDIs $\{P_j\}$ and $\{Q_k\}$, if every $P_j$
commutes with every $Q_k$ these sample spaces are \emph{compatible} and possess
a common refinement consisting of all nonzero products $P_j Q_k$.  In this
situation one can circumvent the single framework rule by using this common
refinement to build a quantum description.  But if the frameworks are not
compatible the corresponding descriptions cannot be combined.  Additional
details of how the SFR governs quantum descriptions and reasoning are found in
Ch.~16 of \cite{Grff02c}.

\subsection{Measurement outcomes}
\label{sbct4.2}

\xb
\outl{Measurement outcomes requires PDI that includes detector states, not
  $[\Psi_3]$ 
}\xa

\xb
\outl{CH uses $\ket{\Psi_3}$ as pre-probability. This follows Born (1926).
}\xa
 
Let us apply the foregoing discussion to the measurement situation shown in
Fig.~\ref{fgr1}. The CH strategy is to first identify the quantum properties,
which is to say the subspaces of the Hilbert space, that should enter the
description.  Here we are interested in a measurement with two possible
outcomes: either detector $a$ has triggered or $b$ has triggered.  While the
individual states $\ket{D^{a*}}\ot \ket{D^b}$ and $\ket{D^a}\ot \ket{D^{b*}}$
can be interpreted as possible outcomes, their macroscopic superposition
$\ket{\Psi_3}$ in \eqref{eqn6} cannot, and indeed the projector $[\Psi_3]$ does
not commute with any of the projectors $[D^{a*}],\,[D^{a}],\, [D^{b*}]$ or
$[D^{b}]$.  (Our notation follows the usual physicist's convention that a
projector $P\ot I$ on a tensor product can be denoted by $P$.)  Consequently,
if we demand that $[\Psi_3]$ be a physical property at time $t_3$, this choice
of framework (i.e., $[\Psi_3]$ and $I-[\Psi_3]$) will prevent us from
discussing the situation in Fig.~\ref{fgr1} \emph{as a measurement}, a physical
process with some specific macroscopic outcome.  Instead we must use a
framework at time $t_3$ that contains the projectors $[D^{a}]$, $[D^{a*}]$,
$[D^{b}]$, $[D^{b*}]$ and their products.  Having made this choice CH employs
$\ket{\Psi_3}$ not as a physical property but as a \emph{pre-probability}, a
mathematical device which can be used to calculate the probability of various
properties via the usual Born formula:
\begin{equation}
 \Pr(D^{a*}) = \mte{\Psi_3}{D^{a*}\ot I} = |\al|^2,\quad
\Pr(D^{b*}) = |\bt|^2,\quad \Pr(D^{a*}, D^{b*}) =0.
\label{eqn8}
\end{equation}
(Here $\Pr(A,B)$ is the probability of the conjunction $A$ \AND\ $B$.)  In
words, the probability is $|\al|^2$ that the $a$ detector has triggered,
$|\bt|^2$ that the $b$ detector has triggered, and 0 that both have triggered.
This way of understanding $\ket{\Psi_3}$ is not a new innovation, as it goes
back to the work of Born in 1926%
\footnote{According to Jammer \cite{Jmmr74}, p.~38, Born's first paper
  \cite{Brn26} proposing the probabilistic interpretation of Schr\"odinger's
  wave arrived at the publishers of {\it Zeitschrift f\"ur Physik} a mere four
  days after Schr\"odinger's concluding contribution (the part containing his
  time-dependent equation) had been sent to the editor of the {\it Annalen der
    Physik}.  In a longer paper \cite{Brn26b} published the same year Born
  stated (p.~804; translation from Jammer, op cit., p.~40): ``The motion of
  particles conforms to the laws of probability, but the probability itself is
  propagated in accordance with the law of causality.''  Thus although Born did
  not use a technical term such as ``pre-probability'' for the wavefunction
  developing unitarily in time, the idea of employing it as a means of
  calculating probabilities was clearly present in his thinking.  One wonders
  how the history of quantum foundations might have been different if the
  physics community, including Schr\"odinger, had taken Born's idea much more
  seriously, and proceeded to build on it.}.%

\xb
\outl{CH resolves 1st measurement problem using appropriate sample space,
  probabilities. Contrast GRW, BM. CH resembles textbook treatment.
}\xa

Thus the first measurement problem is resolved in the CH approach by employing
a sample space that includes the final measurement outcomes (pointer positions)
in its description, together with the unitary time development generated by the
standard Schr\"odinger equation to assign probabilities using Born's formula.
This is in contrast to the spontaneous localization approach which, as noted in
Sec.~\ref{sbct3.2}, employs a stochastic modification of the Schr\"odinger
equation.  And it is quite unlike Bohmian mechanics in that no hidden variables
are added to the Hilbert space description. Instead, CH uses the procedure
students are actually taught in introductory quantum mechanics courses: compute
the unitary evolution of a wave function and then use the result to calculate a
probability (or an average) using Born's rule.  The only difference is that the
textbook may add a qualifying phrase to the effect that what has been
calculated is the probability of something \emph{if it is measured}.  This
qualification is not totally out of place, but it is also not particularly
helpful, and would be unnecessary if the textbook included an appropriate
discussion of quantum properties together with an explanation of how to make a
consistent use of probabilities in the quantum context.  

As noted above, in the CH discussion of measurement outcomes $\ket{\Psi_3}$ is
not regarded as a quantum property, but instead as a pre-probability, a
mathematical device employed to calculate probabilities, and hence no more
``real'' than a probability distribution of the sort one encounters in
classical statistical mechanics.  Indeed, it is perfectly possible to calculate
the probabilities in \eqref{eqn8} in a manner that never makes any reference to
$\ket{\Psi_3}$; see the discussion in Sec.~9.4 of \cite{Grff02c}.  The CH
approach stands in marked contrast with that of the many worlds interpretation,
in which a unitarily evolving wave function, $\ket{\Psi_j}$ in our example, is
regarded as describing what is really going on in the quantum world. But if
$[\Psi_3]$ is a quantum property at time $t_3$ then from the CH perspective
this precludes, invoking the SFR, a simultaneous assignment of pointer
positions, i.e., specific detector states such as $[D^a]$ or $[D^{a*}]$, at
time $t_3$.  Thus CH and the many worlds approach are mutually incompatible.

\section{Consistent Histories and the Second Measurement Problem}
\label{sct5}

\subsection{Families of histories}
\label{sbct5.1}

\xb \outl{Histories; family of histories (framework). Consistency + Born rule
  extension $\ra$ probabilities }\xa

In order to understand how CH resolves the second measurement problem it is
necessary to discuss its approach to the stochastic dynamics of a quantum
system at a succession of times.  It employs a sample space of
\emph{histories}, where a history is a sequence of (quantum) properties at a
succession of times.  For a sequence of times $t_1 < t_2 < \cdots t_n$ a
typical history will have the form
\begin{equation} F_1 \od F_2 \od \cdots F_n,
\label{eqn9}
\end{equation}
where each $F_j$ is a projector corresponding to the quantum property at the
time $t_j$.  There is no assumption that the $F_j$ at successive times are
related by the unitary time development produced by Schr\"odinger's equation.
The symbol $\od$ is a variant of the usual tensor product symbol $\ot$, and
indeed a collection or family of histories, also referred to as a
\emph{framework} or \emph{realm}, can be thought of as represented as a
decomposition of the identity of a \emph{history} Hilbert space consisting of a
tensor product of copies of the Hilbert space used for static properties.  If
such a family refers to events in a \emph{closed} quantum system and satisfies
certain \emph{consistency conditions} it is possible to assign probabilities to
the different histories using Schr\"odinger dynamics and an extension of the
usual Born rule; the reader is referred to Sec.~3.4 of \cite{Grff13} for
additional details.

\xb \outl{$\FC_u$: Unitary time development framework cannot include
  measurement outcomes }\xa

Let us consider the measurement process in Fig.~\ref{fgr1} using a variety of
different history frameworks, beginning with 
\begin{equation}
 \FC_u:\; [\Psi_0]\od\{[\Psi_1],I-[\Psi_1]\}\od\{[\Psi_2],I-[\Psi_2]\}
\od\{[\Psi_3],I-[\Psi_3]\},
\label{eqn10}
\end{equation}
interpreted as follows: At $t=0$ the initial state is $[\Psi_0] =
\dya{\Psi_0}$.  At each later time $t_j$ there is a PDI which in this case is
relatively simple: a choice between $[\Psi_j]$ and $I-[\Psi_j]$, i.e., either
one of these or the other occurs or is true at the time in question. Thus the
$\FC_u$ family consists of $2^3=8$ different histories.  But when
probabilities are assigned to these histories using the extended Born rule,
those containing an $I-[\Psi_j]$ are assigned 0 probability, and hence there
would be no harm in omitting them.  Thus in effect $\FC_u$ contains only a
single history representing unitary time evolution.

\xb
\outl{$\FC_1$ includes final detector states, allows measurement outcomes
}\xa

For reasons already discussed in Sec.~\ref{sbct4.2}, $\FC_u$ cannot represent a
measuring process, as the use of $[\Psi_3]$ at $t_3$ precludes any discussion
of measurement outcomes.  A more satisfactory family from this perspective is
\begin{equation}
 \FC_1:\; [\Psi_0]\od[\Psi_1]\od[\Psi_2]\od
\{[D^{a*}]\ot [D^{b}],\, [D^{a}]\ot [D^{b*}]\},
\label{eqn11}
\end{equation}
where at times $t_1$, $t_2$ and $t_3$ projectors have been omitted if they only
occur in histories having zero probability.  
An application of the extended Born rule assigns probabilities of
$|\al|^2$ and $|\bt|^2$ to the histories terminating with the $a$ and $b$
detectors having detected the particle, in agreement with \eqref{eqn8}. 

\xb
\outl{$\FC_2$ describes particle paths before measurement. History probabilities
}\xa

\xb \outl{Conditional probabilities allow inference of earlier properties from
  measurement outcomes }\xa

In order to describe properties of the particle \emph{before} it was measured
we introduce another family
\begin{equation}
 \FC_2:\; [\Psi_0]\od[\Psi_1]\od\{[\om_2^a],[\om_2^b]\}\od
\{[D^{a*}]\ot [D^{b}],\, [D^{a}]\ot [D^{b*}] \}.
\label{eqn12}
\end{equation}
Note that at $t_2$ $[\om_2^a]$ represents $[\om_2^a]\ot I\ot I$ on the Hilbert
space that includes the spin and the apparatus, and the same convention applies
to $[\om_2^b]$.  Once again properties, such as $I-[\om_2^a]-[\om_2^b]$, which
are assigned zero probability have been omitted. This leaves four histories:
two possibilities at $t_2$ and two at $t_3$.  To these the extended Born rule
assigns probabilities:
\begin{equation}
 \Pr(\om_2^a,D_3^{a*}) = |\al|^2,\quad \Pr(\om_2^a,D_3^{b*})=0;\quad
 \Pr(\om_2^b,D_3^{b*}) = |\bt|^2, \quad \Pr(\om_2^b,D_3^{a*})=0,
\label{eqn13}
\end{equation}
where the subscript 3 has been added to the detector states to indicate the
time at which this property is or is not true.  Combining \eqref{eqn13}
with  \eqref{eqn8} yields the conditional probabilities:
\begin{equation}
 \Pr(\om_2^a\vbB D_3^{a*}) = 1, \quad \Pr(\om_2^b\vbB D_3^{b*})=1.
\label{eqn14}
\end{equation}
Thus if at $t_3$ $D^a$ has triggered one can be certain (conditional
probability 1) that at $t_2$ the particle was earlier on the path $a$ in
Fig.~\ref{fgr1} leading to this detector, or on path $b$ if $D^b$ has
triggered.  An experimental physicist who has designed this type of equipment
will find these conclusions very reasonable.

\xb
\outl{$\FC_3$ includes spin states at $t_1$; values can be inferred from
  measurement outcomes
}\xa

The discussion of particle properties can be extended to a still earlier time
by using the family
\begin{equation}
 \FC_3:\; [\Psi_0]\od\{[z^+],[z^-]\}\od\ I \od
\{[D^{a*}]\ot [D^{b}],\, [D^{a}]\ot [D^{b*}] \},
\label{eqn15}
\end{equation}
where now the spin states appear at $t_1$. Repeating the sort of
analysis applied to $\FC_2$ leads to the conclusion that 
\begin{equation}
 \Pr([z^+]_1\vbB D_3^{a*}) = 1, \quad \Pr([z^-]_1\vbB D_3^{b*})=1,
\label{eqn16}
\end{equation}
where the subscript 1 has been added to $[z^+]$ and $[z^-]$ to emphasize that
these properties refer to time $t_1$, i.e., \emph{before} the particle entered
the magnetic field gradient, see Fig.~\ref{fgr1}. As the apparatus in
Fig.~\ref{fgr1} was designed to measure $S_z$, \eqref{eqn16} confirms that it
operates correctly.

\xb \outl{All $\FC_j$ are good QM, but differ in utility. Different choices
  cannot $\ra$ contradictions }\xa

It is important to note that within CH as a formulation of the fundamental
principles of quantum mechanics the frameworks $\FC_u$, $\FC_1$, $\FC_2$,
$\FC_3$ are all equally valid quantum descriptions, and there is no principle
of quantum physics which demands the use of one rather than another.  What
distinguishes the different frameworks is their utility in describing the
quantum system of interest and drawing physical conclusions about it.  Thus
while $\FC_u$ represents perfectly good physics, it cannot represent a
measurement process when `measurement' is understood in the usual way as a
process resulting in one of a collection of macroscopic outcomes.  Similarly,
it is only $\FC_2$ and $\FC_3$ that can represent a quantum measurement as that
term is employed in experimental physics, where an earlier microscopic property
is inferred from the later macroscopic outcome.  Given the multiplicity of
possible frameworks one might be concerned that different choices could give
rise to contradictory results, but this cannot occur, as shown in Ch.~16 of
\cite{Grff02c}%
\footnote{ Various claims to the contrary, \cite{Knt97}, \cite{BsGh99} and
  \cite{BsGh00} have been responded to in \cite{GrHr98}, \cite{Grff00} and
  \cite{Grff00b}, respectively; short replies by the critics follow immediately
  after each response.}.

\subsection{Quasiclassical frameworks}
\label{sbct5.2}

\xb
\outl{Instead of pure states for apparatus one can use Qcl framework
}\xa

\xb
\outl{Application to measurement process: Ch.~17 of CQT
}\xa

The preceding discussion of the measurement process has been carried out using
only pure quantum states, and the reader may wonder whether this is
appropriate, especially for apparatus states.  Should one not use Hilbert
subspaces of higher dimension, or density operators?  The analysis given here
is not misleading, but it can also be extended to a more general framework in
which both the initial state of the apparatus and its final pointer states have
a \emph{quasiclassical} description. That is to say, one in which those
macroscopic quantities that one expects will be correctly described using
classical mechanics are represented by appropriate Hilbert subspaces of
enormous dimension, in such a way that the stochastic quantum dynamics of the
associated family of histories is in turn well approximated by the
deterministic dynamical laws of classical physics.  While the emergence of
classical mechanics from quantum mechanics in this way has not been rigorously
demonstrated, there is significant work indicating how it can be done.  We
refer the reader to \cite{GMHr93,Hrtl11}, and to Ch.~17 of \cite{Grff02c} for a
specific application to measuring processes.

\xb
\outl{Fundamental QM does not require use of Qcl frameworks
}\xa

In this connection it is important to note that as a statement of fundamental
quantum principles, CH does \emph{not} require the use of quasiclassical
frameworks in situations where they might seem to be a natural choice, as at
time $t_3$ in the Stern Gerlach measurement.  As noted above, the framework
$\FC_u$ is perfectly acceptable as a quantum description, though not very
useful for describing a measurement.  Similarly, in other situations where
quantum mechanics supplies a quasiclassical description there are always other
frameworks which are very far from being quasiclassical, a fact that has
sometimes been used to declare the CH approach unacceptable
\cite{DwKn96,Knt96,Knt98}; see the discussion in Sec.~4.2 of \cite{Grff13}.

\section{Alternative Approaches to the Second Measurement Problem}
\label{sct6}

\subsection{Standard (textbook) quantum mechanics}
\label{sbct6.1}

\xb
\outl{Measurement when M1/M2: Qm system is/is not in estate of measured
observable. M3: Collapse into the measured state
}\xa

Let us begin a discussion of other approaches to the second measurement problem
by considering standard quantum mechanics i.e., what is found in textbooks or
based thereon.  In textbooks a typical discussion of measurements will contain
the following elements:
\begin{description}
\item[M1.]  If a quantum system is initially in an eigenstate of a particular
  observable, a measurement of that observable will with certainty give the
  corresponding eigenvalue.
\item[M2.] If a quantum system is not initially in an eigenstate, then the
  measurement outcome will correspond to one of the observable's eigenvalues,
  with a probability given by the Born formula.
\item[M3.] In case {\bf M2} the measurement will cause the measured system to
jump or collapse into an eigenstate corresponding to the measured eigenvalue.
\end{description}

\xb
\outl{Measurement of eval $a_j$ of observable $A$ $\lra$ property $P_j$ in
  spectral representation of $A$
}\xa

The Hermitian operator $A$ representing a quantum observable has a spectral
representation
\begin{equation}
 A = \sum_j a_j P_j,
\label{eqn17}
\end{equation}
where the $\{P_j\}$ are a PDI, and we assume each eigenvalue appears only once
in the sum: $j\neq k$ implies $a_j\neq a_k$.  Hence $P_j$ projects on the
subspace spanned by all the eigenvectors with eigenvalue $a_j$.  Thus a
measurement of $A$ that yields the value $a_j$ is equivalent to a measurement
of the corresponding PDI that reveals the property $P_j$.

\xb
\outl{CH consistent with M1, M2; M3 will not be discussed  
}\xa

\xb\outl{In general SQM does not allow inferring prior state from measurement
  outcome 
}\xa

The CH discussion in Sec.~\ref{sct4} of the situation represented in
Fig.~\ref{fgr1} is consistent with {\bf M1} and {\bf M2}. (Since {\bf M3} only
applies to the case of a special sort of nondestructive measurement which can be
thought of as combining a preparation with a measurement, see Sec.~\ref{sct2},
it will not be discussed further here; see \cite{Grff14,Grff14b} for a CH
treatment of such measurements.)  If one somehow knows that the system was
initially in some eigenstate of the measured observable, then by {\bf M1} the
measurement outcome can be used to infer this state before the measurement took
place, but for a general superposition state such an inference is not possible.  This is because the textbook approach corresponds to using the family
$\FC_1$ of Sec.~\ref{sbct5.1} when analyzing the measurement in
Fig.~\ref{fgr1}: unitary time development until the final pointer position
emerges.

\xb
\outl{Spherical wave spreading towards detector preceded by collimator
}\xa

\xb
\outl{CH allows framework in which wave packets travel in definite directions
}\xa

From the CH perspective the textbook discussion is not incorrect, but it is
inadequate, and the inadequacy is not entirely trivial, as experimental
physicists often infer something about the preceding trajectory of a particle
from the path revealed by their detectors.  Suppose that a particle is
scattered in an S (spherically symmetrical) wave from a target, and is detected
some distance away by a detector preceded by a collimator, a metal plate with a
small hole in it, that serves to define the precise scattering angle.  When a
particle is detected the experimenter makes the seemingly reasonable inference
that it followed a straight line from some point on the target and passed
through the hole in the collimator on its way to the detector. The CH approach
allows the use of a framework in which the particle emerges from the scattering
center in a spherical wave, but also a different framework, incompatible with
the first one, with histories in which the particle travels outwards from the
target as a wavepacket in a definite (but not infinitely precise) direction.
This second framework justifies the experimenter's intuition and resolves one
of the standard paradoxes of quantum mechanics first pointed out by Einstein
(see pp.~115f in \cite{Jmmr74}), the seemingly instantaneous ``collapse'' of a
spherical wave upon detection.

\xb
\outl{Students despite textbooks reach reasonable conclusions
}\xa

\xb
\outl{CH clarifies stuff in textbooks; is ``Copenhagen done right''
}\xa

Despite the fact that textbooks do not provide an adequate discussion of how to
deduce prior states from measurement outcomes, students trained by this
approach, with its emphasis on calculational techniques rather than genuine
physical understanding, are nonetheless often able to draw reasonable
conclusions when they think about what is going on in a particle experiment,
such as the one described in the preceding paragraph, conclusions which can be
justified by CH methods.  Thus CH can be regarded as clarifying rather than
being in direct conflict with much of what is found in textbooks, justifying
the claim of its proponents that instead of being some completely new theory
CH is ``Copenhagen done right.''

\subsection{Spontaneous localization and many worlds}
\label{sbct6.2}

\xb
\outl{Both GRW and BM resolve first measurement problem; for MW it is disputed
}\xa

\xb
\outl{Second problem not resolved by GRW or MW
}\xa

The spontaneous localization and many worlds approaches to the first
measurement problem were discussed above in Sec.~\ref{sbct4.2}.  Neither
approach has a satisfactory answer to the the second measurement problem, for
reasons which can be readily understood with reference to Fig.~\ref{fgr1}.  Let
us begin with spontaneous localization, which replaces the Schr\"odinger
equation with a stochastic variant that causes the wavefunction to
spontaneously collapse in the course of time.  The collapse rate is chosen to
be very slow so as not to be in obvious disagreement with existing experiments,
but as a consequence when the particle emerges from the field gradient at time
$t_2$ in Fig.~\ref{fgr1} it will be in (or very close to) a coherent
superposition of the position wave packets $\ket{\om_2^a}$ and $\ket{\om_2^b}$.
Consequently, from the later result that, say, $D^a$ triggered, there is no way
of inferring that the particle at $t_2$ was in the upper track in
Fig.~\ref{fgr1}, much less the property $S_z=+1/2$ at $t_1$.  Thus this
formalism does not contain the mathematical machinery needed to support an
inference which experimental physicists make very easily: the detector
triggered because a particle was traveling towards it.  The same criticism
applies to the many worlds approach for which there is nothing at any time
except for the unitarily developing Schr\"odinger wave. Obviously nothing in
$\ket{\Psi_j}$ for $j=0, 1, 2$ gives any hint as to which detector will later
detect the particle, and thus many worlds lacks a mathematical procedure for
making the required inference.  It is hard to see how ``splitting universes''
can resolve this problem unless the splitting occurs by the time the wave
packet emerges from the magnetic field gradient.  And that does not seem very
satisfactory given that experiments exist for detecting the interference of
matter waves spread out over small but still macroscopic distances.

\subsection{Bohmian mechanics}
\label{sbct6.3}
\xb
\outl{2d measurement problem. BM is OK for Fig. 1, but not for crossed beams
discussed by Bell
}\xa

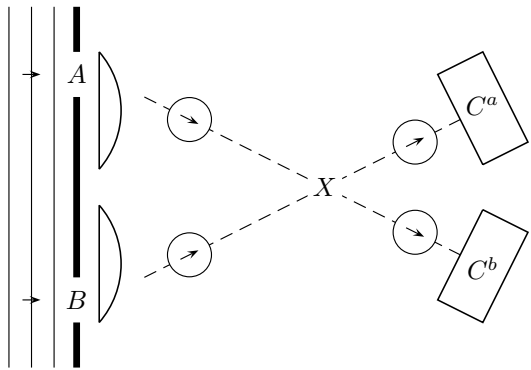
\begin{figure}[h]
$$
\begin{pspicture}(-1.8,-2.4)(5.4,2.4) 
\newpsobject{showgrid}{psgrid}{subgriddiv=1,griddots=10,gridlabels=6pt}
\def\lwd{0.035} 
\def\lwb{0.15}  
\def\lwn{0.025}  
\psset{unit=0.6cm, 
labelsep=2.0,
arrowsize=0.150 1,linewidth=\lwd}
\def\brecs(#1,#2){%
\psframe[fillcolor=white,fillstyle=solid,linestyle=none](-#1,-#2)(#1,#2)}
\def\wup{\pscircle[linewidth=\lwn,fillcolor=white,fillstyle=solid](0,0){.5}
\psline[linewidth=\lwn]{->}(-0.2,-0.1)(0.2,0.1)}
\def\wdn{\pscircle[linewidth=\lwn,fillcolor=white,fillstyle=solid](0,0){.5}
\psline[linewidth=\lwn]{->}(-0.2,+0.1)(0.2,-0.1)}
\psline[linewidth=\lwb](-1.5,-4)(-1.5,-3)
\psline[linewidth=\lwb](-1.5,-2)(-1.5,+2)
\psline[linewidth=\lwb](-1.5,+4)(-1.5,+3)
\psline[linewidth=\lwn](-2,-4)(-2,4)
\psline[linewidth=\lwn](-2.5,-4)(-2.5,4)
\psline[linewidth=\lwn](-3,-4)(-3,4)
\psline[linewidth=\lwn]{->}(-2.7,-2.5)(-2.3,-2.5)
\psline[linewidth=\lwn]{->}(-2.7,+2.5)(-2.3,+2.5)
\psarc(-2.5,-1.7){1.985}{-40.9}{40.9}
\psline(-1,-3)(-1,-.4)
\psarc(-2.5,+1.7){1.985}{-40.9}{40.9}
\psline(-1,+3)(-1,+.4)
\psline[linewidth=\lwn,linestyle=dashed](0,-2)(7,1.5)
\psline[linewidth=\lwn,linestyle=dashed](0,+2)(7,-1.5)
\rput(1,-1.5){\wup}\rput(6,1){\wup}
\rput(1,+1.5){\wdn}\rput(6,-1){\wdn}
\psline(7.5,0.5)(8.5,1)(7.5,3)(6.5,2.5)(7.5,0.5)
\psline(7.5,-0.5)(8.5,-1)(7.5,-3)(6.5,-2.5)(7.5,-0.5)
\rput(-1.5,2.5){$A$}
\rput(-1.5,-2.5){$B$}
\rput(4,0){\brecs(0.4,0.6)}
\rput(4,0){$X$}
\rput(7.5,1.8){$C^a$}
\rput(7.5,-1.8){$C^b$}
\end{pspicture}
$$
\caption{Double slit interference with bent beams that cross at $X$
and are later detected by $C^a$ and $C^b$. The circles represent wave packets
before and after they cross at $X$. 
}
\label{fgr2}
\end{figure}

The situation is different for Bohmian mechanics because in this interpretation
a particle has a well-defined trajectory in real space, and using this
trajectory makes it possible to infer from detection by $D^a$ that the particle
was at time $t_2$ on the upper $a$ track in Fig.~\ref{fgr1}.  However, there is
another situation in which Bohmian mechanics gives an answer that does not
agree with physical intuition. It is discussed by Bell, see Ch.~14 of
\cite{Bll87}, using a gedanken experiment inspired by Wheeler \cite{Whlr78}, a
variant of the well-known double slit situation shown here schematically in
Fig.~\ref{fgr2}.  The Schr\"odinger wave for a single particle arrives from the
left as a plane wave, and the part of the wave that passes through the $AB$
slit system is refracted by passive elements so that the resulting beams are
bent and cross each other before traveling on to the two detectors $C^a$ and
$C^b$.  The $X$ in the figure labels the position in the vacuum where the beams
cross; there is nothing at this point for the particle to interact with. Unlike
the wave, the Bohmian particle has a precise position at every time and follows
a continuous trajectory in space; which trajectory it follows depends on its
initial position, which is random. Some trajectories pass through the upper and
some through the lower slit.  Assume the particle is eventually detected by
detector $C^a$. Did it earlier pass through the upper slit $A$ or the lower
slit $B$?  When asked this question a physicist is likely to reply $B$, an
answer which Bell dismisses as resulting from a ``naive classical picture.''
The answer given by Bohmian mechanics is $A$, because the quantum wave function
is symmetric under a reflection that interchanges the two slits, and this means
the Bohmian particle trajectory can never cross the symmetry plane midway
between the slits, from top to bottom or vice versa.  Indeed, an analysis of
Bohmian particle trajectories shows that those that come through slit $A$ and
descend downwards towards $X$ eventually ``bounce'' back upwards towards $C^a$,
while those coming through $B$ bounce downwards and eventually reach $C^b$.

\xb
\outl{CH result for crossed beams: particle does not bounce
}\xa

\xb
\outl{Englert ao:  detector triggered when BM particle far away
}\xa

\xb \outl{CH: detector triggers iff particle goes thru it }\xa

\xb \outl{Thus BM answer to 2d measurement problem is not reliable }\xa

However, the CH analysis \cite{Grff99b} of this situation, which is fully
quantum mechanical, when addressing the question of which slit a particle
detected by $C^a$ had earlier passed through gives the answer $B$, contrary to
Bohmian mechanics, but in agreement with naive classical thinking.  The
particle passes through $X$ without a bounce. In addition, a numerical study by
Englert et al. \cite{ESSW92} showed that if a certain type of nondestructive
detector $N$ (not shown in Fig.~\ref{fgr2} ) that is triggered by a particle
passing through it is placed on the lower path between $B$ and $X$, it
registers the particle's passage in coincidence with $C^a$---either both are
triggered, or neither.  However, when this situation is analyzed using Bohmian
mechanics, it is sometimes the case that the Bohmian particle follows the upper
path on its way to $C^a$, and nonetheless the detector $N$ is triggered.  From
this Englert et al.\ concluded that the Bohmian particle trajectory is
``surrealistic'', without physical significance. That Bohmian trajectories can
have the strange behavior pointed out in \cite{ESSW92} was later confirmed
through careful numerical calculations by Dewdney et al. \cite{DwHS93} who,
unlike the authors of \cite{ESSW92}, were in favor of, or at least not critics
of, Bohmian mechanics.  The supporters of Bohmian mechanics have defended their
claim that the bouncing particle trajectory is physically plausible by, among
other things, asserting that standard quantum mechanics cannot say anything
about the position of a particle prior to its measurement, or is at best
ambiguous \cite{DFGZ93,Hlln15}. But the CH analysis in \cite{Grff99b}, which
extends standard quantum mechanics in an unambiguous way that resolves the
second measurement problem, indicates that $N$ will be triggered if and only if
the particle passes through it, a result which most physicists would consider
reasonable.  For further details the reader is referred to \cite{Grff99b} and
the references given there.  The conclusion is that the answer Bohmian
mechanics gives to the second measurement problem is not always credible.

\section{The Criticisms of Okon and Sudarsky}
\label{sct7}

\xb
\outl{After discussing GRW, BM we reply to OS criticisms of CH approach to 
measurements
}\xa

Now that the application of the CH approach to a particular measurement
situation has been discussed in some detail and compared with certain other
quantum interpretations, including spontaneous localization and Bohmian
mechanics, both of which Okon and Sudarsky consider superior to CH, we are in a
position to address the criticisms found in \cite{OkSd14b}.  Mainly these
concern how CH approaches the first measurement problem, with some
additional remarks about how it resolves the second problem.


\subsection{First measurement problem}
\label{sbct7.1}

\xb \outl{Quotations from O\&S paper: CH relies on elements outside its
  formalism. It cannot specify correct framework for obtaining reliable
  predictions }\xa

The principal criticism in \cite{OkSd14b} of the CH approach to the first measurement
problem will be evident from the following quotations. (Here an `a' and `b'
following a page number indicate the first and second column, respectively.)

\begin{description}
\item[Q1.]
We find [the CH treatment of measurements] unsatisfactory because it relies,
often implicitly, on elements external to those provided by the formalism. In
particular, we note that, in order for the formalism to be informative when
dealing with measurement scenarios, one needs to assume that the appropriate
choice of framework is such that apparatuses are always in states of well
defined pointer positions after measurements. The problem is that there is
nothing in the formalism to justify this assumption. (Abstract)
\item[Q2.] 
  [A]mong all the possible frameworks, only one is suitable to describe what in
  fact we perceive or experience\dots [T]he main problem is that CH is
  incapable of recognizing in advance, and without bringing in elements that
  rely on our intuition and experience, which is going to be the framework that
  does the job. We conclude, then, that CH does not really constitute a
  satisfactory solution to the measurement problem. (Sec.~3, p.~9b)
\item[Q3.] 
  [T]he fact that a given measuring apparatus actually measures some property
  is something that cannot be deduced from the CH formalism but that must be
  discovered by experience.  That is, if one is given a new measurement
  equipment, described entirely in quantum terms (via a Hamiltonian, an initial
  state, etc.), CH is unable to answer, unlike, say [spontaneous localization]
  or Bohmian mechanics, which are going to be its possible final states when we
  use it in the laboratory.  Therefore, in a given situation, described by
  providing the complete physical set-up in terms of the initial state of the
  closed system and the Hamiltonian, CH is incapable of predicting which
  framework one must choose. (Sec.~3, p.~10b)
\item[Q4.] 
  [T]he truly problematic point\dots is that the CH formalism is incapable of
  picking out the right framework. That is, it does not offer any clear
  characterization regarding the exact relation between the experimental set-up
  and the framework one needs to use in order to get reliable predictions out
  of the theory. (Sec.~3, p.~11a)
\end{description}

\def\qa{\textbf{Q1}}
\def\qb{\textbf{Q2}}
\def\qc{\textbf{Q3}}
\def\qd{\textbf{Q4}}
\def\qe{\textbf{Q5}}

\xb
\outl{Q1. `Measurement' properly excluded from fundamental CH principles;
need additional stuff to apply these to measurement.
}\xa

\xb
\outl{Analogy: `Telescope' not among fundamental principles of optics
}\xa

In response to the assertion in \qa\ that CH employs elements external to the
formalism when discussing measurements, we note that this is quite proper
insofar as CH is a statement of the fundamental principles of quantum
mechanics.  Indeed, most physicists working in quantum foundations do not think
that `measurement' should be included among the fundamental principles, and
surely one of the main motivations behind the spontaneous localization, Bohmian
mechanics, and many worlds approaches is to find principles that do \emph{not}
refer to measurements.  To use an analogy, one can employ the fundamental
principles of optics, such as refraction and reflection, to describe the
operation of a telescope, but it would be extremely odd to find `telescope'
playing an essential role among the principles of optics, which apply equally
to microscopes, eyeglasses, the human eye, and so on.  Thus in the case of
quantum measurements one would anticipate that any plausible fundamental theory
of quantum mechanics would contain no reference to them, and therefore
additional concepts must be employed in order to to describe them.  The
proper question is not whether elements external to the formalism are
used, but instead whether these ``extras'' are needed because one is
discussing a measurement and not some other physical process.

\xb \outl{Macroscopic outcomes (quasiclassical framework) a necessary part of
  `measurement' }\xa

\xb
\outl{CH approach different from GRW and BM, but this does not mean it's wrong
}\xa

To be specific, the complaint in \qa\ is that CH must choose a framework in
which the apparatus has suitable outcome states, pointer positions, at the end
of the measurement.  That is correct: in the CH approach all four families of
histories introduced in Sec.~\ref{sbct5.1} represent valid quantum descriptions
of the situation shown in Fig.~\ref{fgr1}, but only the last three include the
pointer positions, and hence constitute possible descriptions of a measurement
understood as a physical process in which prior microscopic properties are
later revealed or correlated with macroscopic outcomes.  A framework that makes
no reference to the microscopic properties or to the macroscopic outcomes
cannot describe a 'measurement' as that term is understood in experimental
quantum physics, any more than a discussion omitting all reference to its
mirror and the image it produces could be said to describe a reflecting
telescope.  Since macroscopic properties are described in the CH approach using
a quasiclassical framework, Sec.~\ref{sbct5.2}, and the superposition state
$\ket{\Psi_3}$ does not belong to such a framework, it cannot possibly be used
when describing a measurement, a point nowhere better stated than in Bell
\cite{Bll90}.  In thus excluding $\FC_u$ from the collection of frameworks
which might describe a measurement, CH is employing the single framework rule,
which is very much part of its formalism, along with the notion of a
quasiclassical framework which, although not a fundamental quantum principle,
is what it uses to describe the macroscopic world.
To be sure, this is very different from the paths followed by spontaneous
localization (modified Schr\"odinger dynamics) and Bohmian mechanics (use of
particle positions and dynamics in addition to the Hilbert space) in their
approach to the first measurement problem.  But that it is different does not
mean it is wrong or inadequate or inferior or incomplete, as \qc\ would seem to
imply.

\xb
\outl{Q2. What humans perceive is (plausibly) given by Qcl framework
}\xa

Turning to \qb, the topic of what human beings can perceive and experience is a
fairly subtle one, and from the quantum perspective not free from controversy.
However, if one is willing to allow that those perceptions and experiences most
relevant to the present discussion can be described using a quasiclassical
framework, then the response can be similar to that given above for \qa: only
quasiclassical properties and processes will be of use in a quantum description
of something people can perceive, and this fact can again be used to rule out
$\FC_u$, with its far-from-quasiclassical $\ket{\Psi_3}$, as appropriate for
describing a measurement whose outcome can be seen and discussed by scientists.

\xb
\outl{Q3. Unlike GRW, BM, CH allows different sample spaces; the fact
that apparatus measures some property constrains the choice
}\xa

What seems really at stake in \qc\ is the fact that spontaneous localization
and Bohmian mechanics, both of which are stochastic, employ sample spaces which
are fixed in advance by the postulates of the theory, whereas CH does not
specify the sample space in advance: each of the frameworks introduced in
Sec.~\ref{sbct5.1} is a separate sample space, and they are mutually
incompatible so they cannot be combined.  This structure is certainly distinct
from anything one finds in classical physics, and the final statement in \qc,
that the CH approach is ``incapable of predicting which framework one must
choose'' is correct.  The choice of a framework is one made by the physicist
constructing a quantum description, and there is no constraint among the
principles of CH that prescribes which one must be used.  However, the
consistent historian will add that if the ``given measuring apparatus actually
measures some property,'' that fact alone is enough to constrain the choice of
an appropriate framework to one such as $\FC_2$ or $\FC_3$ among those
discussed in Sec.~\ref{sbct5.1}, contrary to the claim in the final sentence in
\qc.  And it is this kind of ``utilitarian'' consideration that allows CH to
solve the second measurement problem, one which, see Sec.~\ref{sct6}, 
standard quantum mechanics, spontaneous localization, Bohmian mechanics, and
many worlds are unable to address.

\xb
\outl{Q4. See replies to Q1-Q3. Also: notion of 'right' framework comes from
the Cl world. Unicity is likely central to OS crits, is not valid in Qm world
}\xa

As for \qd\, the discussion given in Secs.~\ref{sct4} and \ref{sct5},
together with the comments above about \qa, \qb, and \qc, should constitute a
sufficient answer to the second sentence, and need not be repeated.  But the
first sentence, the assertion that ``the CH formalism is incapable of picking
out the right framework'' deserves additional comment.
The notion that there must be a single ``right'' framework comes from the
classical world, from classical physics, where it is assumed, often implicitly,
that at any given time there is a single true state of the world. In Sec.~27.3
of \cite{Grff02c} this notion is given a technical name: \emph{unicity}.  From a
physicist's perspective unicity is exemplified by the fact that in a classical
phase space a single point represents the exact and true state of the system at
a particular time; all properties of the system represented by sets of points
that include this point are true, and those that do not include this point are
false.  By contrast, in a Hilbert space there is no single subspace with a
comparable property. Hence if a Hilbert space, rather than a classical phase
space or a classical collection of hidden variables, provides the correct
mathematical description of physical reality, one might expect the principle of
unicity to fail, and thus in an absolute sense there is no ``right framework.''

\subsection{Second measurement problem}
\label{sbct7.2}

\xb \outl{O\&S criticism Q5. Many frameworks possible; why take any one
  seriously?  Also, d'Espagnat has given a stronger argument against the CH
  approach.  }\xa

The criticism in \cite{OkSd14b} of the CH approach to the second measurement
problem is relatively brief.  After citing a particular example from
\cite{Grff02c} the authors comment that:
\begin{description}
\item[Q5.]  It seems\dots that one can say that some particle had a definite
  property before a measurement took place only if one chooses a specific
  framework where this is so. But, if according to CH, all frameworks are
  equally valid, it is not clear why the description according to this one
  framework should be taken that seriously.  That is, the problem is solved in
  only one of an infinite number of possible frameworks.
\end{description}
Following this is a footnote in which they claim that d'Espagnat \cite{dEsp87}
in 1987 provided a stronger argument against the possibility of CH solving the
second measurement problem.

\xb
\outl{Q5 reply. Result from one framework not invalidated by other frameworks.
Measurement of a particular property requires a framework that contains it. 
}\xa

Let us begin with \qe.  It is indeed the case that one can only discuss the
property a particle had before the measurement by choosing a specific framework
that includes the property in question at the appropriate time.  Thus it is
necessary to use $\FC_3$ in Sec.~\ref{sbct5.1}, or some other framework which
includes the same possibilities at $t_1$, in order to discuss whether the
measurement outcome tells one the value of $S_z$ before the measurement took
place.  This should not be surprising; in the CH approach quantum properties
can only be discussed when they appear in a suitable framework, one that
contains the appropriate projectors as part of the event algebra.  We have
already discussed this in some detail with reference to measurement outcomes in
Sec.~\ref{sbct7.1}, and exactly the same principle applies to measured
properties.  The assertion in \qe\ that because there are alternative
frameworks the result obtained using the $S_z$ framework ``should not be taken
seriously'' seems extremely odd.  Is a news account stating that it was snowing
on a particular day in Pittsburgh rendered invalid by an alternative news
account which describes actions by the city council on the same day, but makes
no mention of snow?  Were it the case that different quantum frameworks gave
contradictory results, this would indeed be a cause for concern.  But they do
not; the consistency of CH inferences is guaranteed by the considerations given
in Ch.~16 of \cite{Grff02c}.  Thus the existence of other frameworks is
entirely beside the point, as it no way invalidates the description obtained
using a particular one.  There are also an infinite number of possible
frameworks for discussing the situation at time $t_3$ in Fig.~\ref{fgr1}, but
most of them do not allow any discussion of the measurement outcome as having
definite results, without which there is no reason to call it a measurement.
Similarly, choosing a framework in which the quantum properties the measurement
was designed to measure cannot even be discussed disqualifies using the term
``measurement'' as that is normally employed in experimental physics, and as it
ought to be employed in discussions of quantum foundations.

\xb
\outl{d'Espagnat had 2 arguments. First violated SFR. Second analogous to
``delayed choice'' and involves counterfactuals; discussed in CQT. Recent
RBG-Stapp exchange
}\xa

The 1987 paper by d'Espagnat \cite{dEsp87} contains two arguments which might
be considered as counting against the CH solution of the second measurement
problem. It focuses on the property of a spin-half particle at a time $t_1$
between when it is prepared at time $t_0$ and when it is measured at time
$t_2$.  The first argument requires combining two incompatible frameworks in a
manner forbidden by the single framework rule, which had not been as clearly
formulated in 1987 as it was later.  For an extended response to this and
similar criticisms by d'Espagnat see Sec.~V~B of \cite{Grff98}. The second
argument involves what would happen if a different measurement were carried out
at time $t_2$, and can be seen as a version of Wheeler's ``delayed choice''
experiment \cite{Whlr78} involving counterfactual considerations: something was
measured but something else could have been measured.  Counterfactuals and
delayed choice experiments are discussed in considerable detail in Chs.~19 and
20 of \cite{Grff02c}, where it is shown that the CH treatment is both
consistent and coherent.  For a more recent discussion involving
counterfactuals see the exchange between Stapp and the author
\cite{Stpp12,Grff12b}.

\section{Conclusion}
\label{sct8}

\xb \outl{In \S\S 4, 5 CH resolves both measurement problems (as define in \S
  2) for the SG expt of \S 3, using frameworks, Born rule, Schr dynamics}\xa

\xb
\outl{CH innovation: consistent probabilities, multiple frameworks, SFR
}\xa

With reference to a particular Stern Gerlach measurement situation introduced
in Sec.~\ref{sct3} we have shown in Secs.~\ref{sct4} and \ref{sct5} how the
consistent histories (CH) approach resolves both parts of the quantum
measurement problem as defined in Sec.~\ref{sct2}, allowing a 
measurement, a macroscopic output revealing a prior microscopic quantum
property, to be discussed from beginning to end in fully quantum mechanical
terms.  
This requires the introduction of appropriate frameworks or quantum sample
spaces, and the use of the extended Born rule to assign probabilities with the
help of (unmodified) Schr\"odinger dynamics.  The main innovation of the CH
approach relative to standard (textbook) quantum mechanics is the use of a
fully consistent system of probabilities at both the microscopic and
macroscopic levels. This allows a quantum system to be described using a
variety of different frameworks, but then insists (the single framework rule)
that incompatible frameworks cannot be combined.

\ca
 a multitude of incompatible (in
the quantum sense) frameworks along with the single framework rule
The CH approach allows the stochastic time development of the measurement
process to be described by a variety of frameworks, which is to say consistent
families of histories, to which probabilities are assigned using the extended
Born rule along with (unmodified) Schr\"odinger dynamics.
\cb

\xb
\outl{Only some frameworks describe a \emph{measurement} as such
}\xa

\xb
\outl{Qm foundations emphasis on 1st measurement problem has led to neglect of
  2d.  Both must be resolved
}\xa

Four of a large number of possible frameworks for the measurement situation
shown in Fig.~\ref{fgr1} and described in Sec.~\ref{sct3} are discussed in
Sec.~\ref{sbct5.1}, and they illustrate the fact that only certain frameworks
are suitable for discussing this measurement \emph{as a measurement}, using
that term in the way it is employed by physicists who conduct actual
experiments.  In particular, an acceptable framework should include both
microscopic properties before the measurement takes place and macroscopic
outcomes at its conclusion.  It is much to be regretted that the longstanding
confusion surrounding the measurement problem in quantum foundations has led to
a focus on the problem of understanding measurement outcomes in quasiclassical
terms (getting rid of Schr\"odinger's cat), the \emph{first} measurement
problem in the terminology used in this paper, to the neglect of addressing the
equally important question of how this outcome is related to the microscopic
property that was (supposedly) measured, in our terminology the \emph{second}
measurement problem.  Both problems must be resolved if one wants to claim that
experimental results of measurements confirm the correctness of quantum theory.
At present CH appears to be the only quantum interpretation that resolves the
second problem.

\xb
\outl{Main OS claim: framework choice for measurement outcome requires going
  outside CH formalism; inferior to BM and GRW
}\xa

\xb \outl{Reply: CH only adds to its formalism features specific to measurement
}\xa

The principal criticisms of Okon and Sudarsky \cite{OkSd14b}, see
Sec.~\ref{sbct7.1}, when applied to the example in Sec.~\ref{sct3} as analyzed
from the CH perspective in Secs.~\ref{sct4} and \ref{sct5}, has to do with
choosing one of the frameworks $\FC_1$, $\FC_2$, or $\FC_3$ in
Sec.~\ref{sbct5.1}, rather than $\FC_u$, in order to describe the measurement
outcome.  These authors consider the CH approach, which rejects a family of
histories containing $\ket{\Psi_3}$ at time $t_3$ because it cannot represent a
quantum property corresponding to a macroscopic measurement outcome, as
inadequate, or at least inferior to the spontaneous localization and Bohmian
mechanics procedures for resolving the first measurement problem.  Our
response, given in detail in Sec.~\ref{sbct7.1}, is that the CH approach
employs its own principles correctly, and does not go outside of its formalism
except to include features specific to a measurement as that term is understood
by physicists, and hence necessary for this particular application of these
principles.  As a fundamental theory of quantum mechanics CH makes no
reference to measurements, which is one reason it can be considered an
advance over standard (textbook) quantum theory. 

\xb
\outl{OS criticism of CH solution to 2d measurement problem: too many frameworks
}\xa

\xb
\outl{Reply: Qm description of measurement must include prior property
}\xa

\xb
\outl{GRW, MW, cannot solve 2d measurement problem. BM gives the wrong answer
}\xa

These authors also criticize the CH solution to the second measurement problem,
Sec.~\ref{sbct7.2}, again on the grounds that there is no way within the
formalism of identifying the \emph{correct} framework.  Our response, as
before, is that if a measurement is supposed to have measured some microscopic
quantum property, the quantum description of the measurement process must
necessarily include that property at a time before the measurement took place,
and this suffices for identifying an appropriate framework.  As noted in
Sec.~\ref{sbct6.2}, it seems clear that neither the spontaneous localization
nor the many worlds interpretations possess the means to resolve the second
measurement problem, whereas Bohmian mechanics does not always give the right
answer, i.e., a solution that working physicists would consider plausible.  In
addition, comments on some out-of-date criticisms of CH by d'Espagnat have been
placed at the end of Sec.~\ref{sbct7.2}.

\xb
\outl{Unicity probably basic to OS criticisms. Is modern counterpart of
  immobile earth
}\xa

\xb \outl{Unicity basic to GRW, BM, MW; is why they can't solve  2d
  measurement problem }\xa

Although there is no discussion of it in \cite{OkSd14b}, it seems likely that
fundamental to the Okon and Sudarsky criticism is their adherence to the
concept of unicity, the idea that at any single time the world is described by
precisely one mechanical state.  As noted in Sec.~\ref{sbct7.2}, this idea fits
very well with classical mechanics, where a single point in its phase space
provides a complete characterization of a mechanical system, but is
incompatible with the structure of Hilbert space.  The fact that from the
quantum perspective a single quasiclassical framework suffices for all of
classical physics provides a plausible explanation of why the concept of
unicity is so deeply embedded in human thinking, and thus why it is so
difficult to accept the notion that at the quantum level there are multiple
incompatible frameworks and unicity no longer holds.  If the CH approach is
correct then, as noted in Sec.~\ref{sct1}, unicity in the modern
context is somewhat analogous to the notion of an immobile earth, which had to
be discarded with the rise of modern cosmology, and thus abandoning unicity is
necessary in order to reach a proper understanding of the quantum world.
Unicity is also maintained in the spontaneous localization, Bohmian mechanics,
and many worlds interpretations, and it is what separates them from CH at the
most fundamental level.  

\xb
\section*{Acknowledgments}
\xa

I thank F.~Lalo\"e and P.~Pearle for helpful correspondence. The research
described here received support from the National Science Foundation through
Grant 1068331.

\end{document}